\documentclass[RNAAS]{aastex701}

\begin{document}

\title{Systematic Effects in Gaia CRF Photometry Related to Scanning Geometry and BP/RP excess factor}

\author[0000-0003-2336-7887]{Valeri V. Makarov}
\affiliation{U.S. Naval Observatory, 3450 Massachusetts Ave NW, Washington, DC 20392-5420, USA}
\email{valeri.makarov@gmail.com}


%%
%% RNAAS manuscripts do not have abstracts.
%% The first paragraph of the body serves as a brief summary.
%%
\begin{abstract}
I report a previously undocumented, large-amplitude systematic dependence
of the median Gaia $G$-band magnitude of CRF3 sources on \texttt{ra\_dec\_corr}, which reflects the preferred direction of scans,
accompanied by a partially correlated trend in the BP/RP flux excess factor. Given the
imminent release of Gaia DR4, the purpose of this note is alerting the community to this artifact.

\end{abstract}
\section{Introduction}

The third realisation of the Gaia Celestial Reference Frame
\citep[Gaia-CRF3;][]{GaiaCRF3_2022} defines the optical counterpart of the
International Celestial Reference System at sub-milliarcsecond precision.
It comprises approximately 1.6 million QSO-like sources drawn from the
\textit{Gaia} Early Data Release 3 \citep[EDR3;][]{GaiaEDR3_2021} catalogue,
of which about 1.2 million carry five-parameter astrometric solutions.
Systematic errors in the broad-band photometry of EDR3 are described in
\citet{Riello2021}, while scanning-law-induced dependencies of the astrometric
covariance structure are discussed in \citet{Lindegren2021_astrom}.

The astrometric catalogue provides the full covariance matrix of each source's
fitted parameters. The correlation between right ascension and declination,
\texttt{ra\_dec\_corr} $\equiv \rho(\alpha,\delta)$, encodes the scan-angle
geometry at the source's sky position averaged over the specific time cadence of measurements. As established by \citet{Lindegren2021_astrom},
these coefficients are determined primarily by the distribution of scan directions
governed by the scanning law. The scanning law is approximately symmetric
with respect to the ecliptic, so $\rho(\alpha,\delta)$ varies systematically
with ecliptic latitude: near-zero values tend to occur where scan directions
are well-distributed in position angle (broadly high ecliptic
latitudes), while large positive or negative values occur where scanning
directions are more degenerate (the ecliptic belt).

\section{Results}

I use the five-parameter solution subset of Gaia-CRF3, queried directly from
the Gaia Archive. The sample includes 1.215 million CRF sources (quasars and AGNs) with 5-parameter solutions. All analyses use binned median statistics to suppress the effect of outliers. Bins of $\sim$12,000 sources are used throughout.

\subsection{G magnitude versus \texttt{ra\_dec\_corr}}

\begin{figure}[ht!]
\includegraphics[width=.45\textwidth]{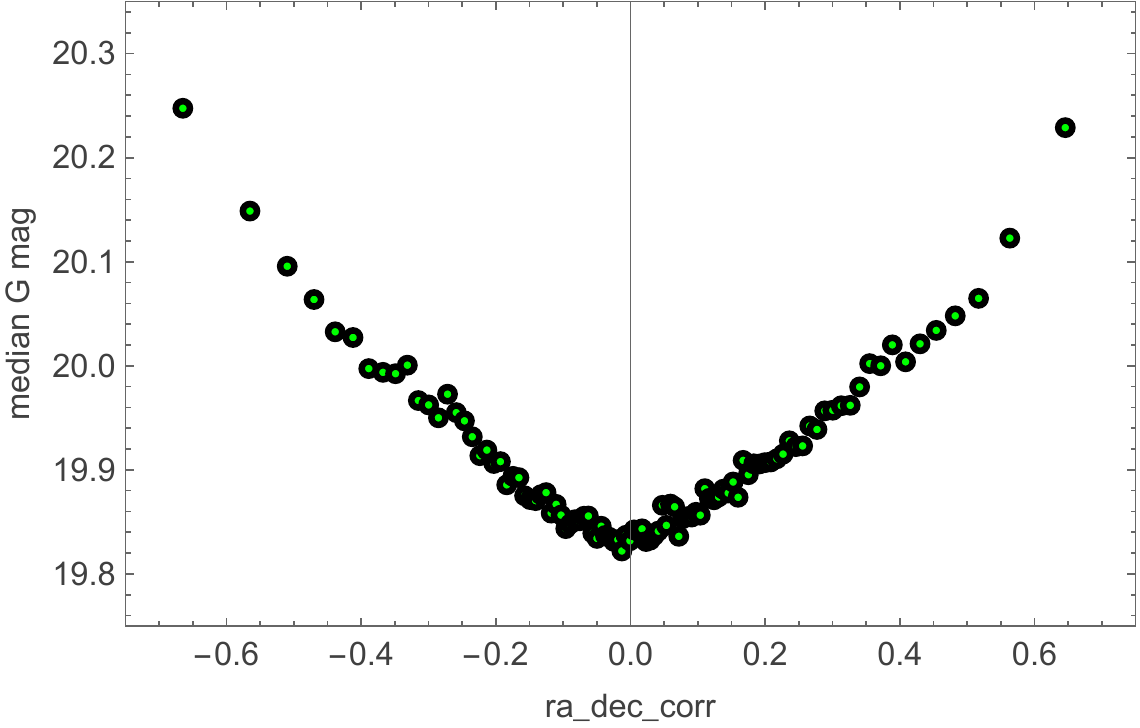}
\includegraphics[width=.45\textwidth]{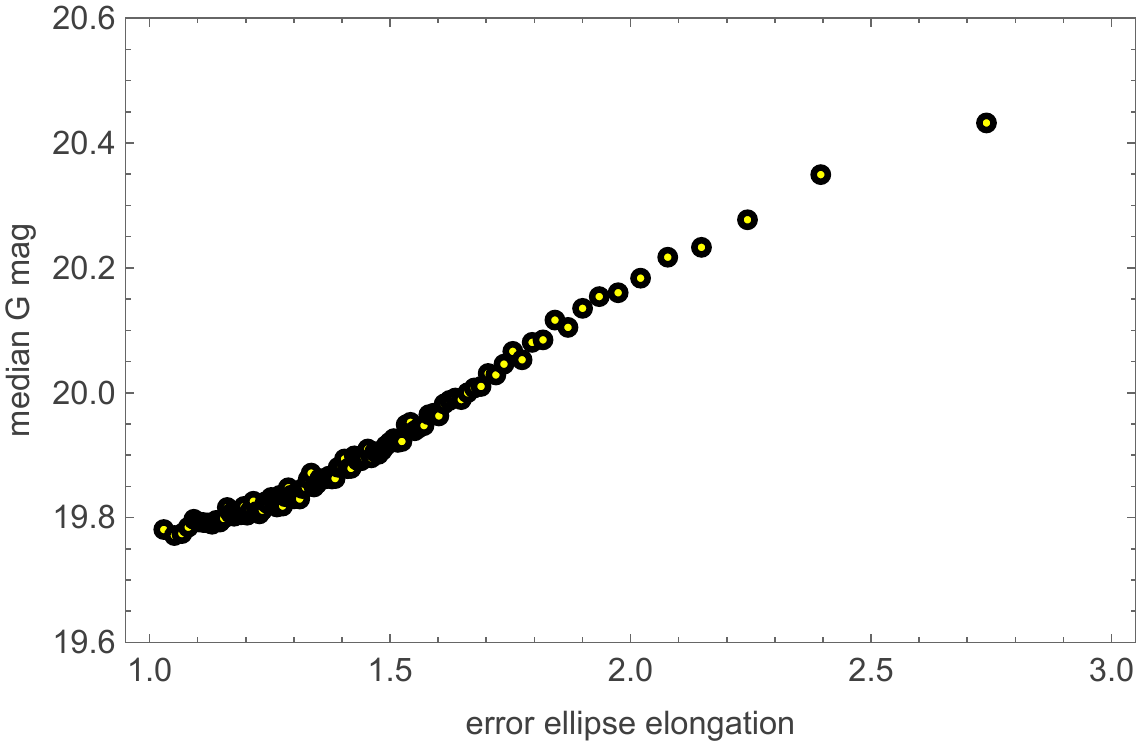}
\includegraphics[width=.45\textwidth]{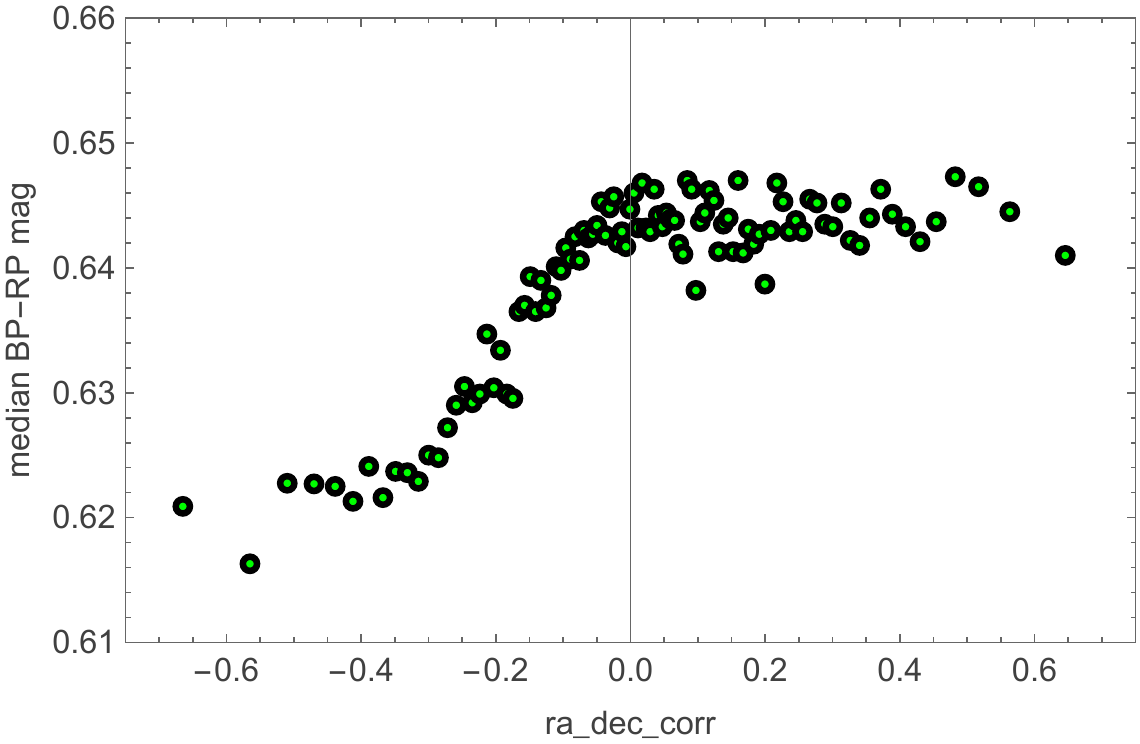}
\hspace{1.5cm}
\includegraphics[width=.45\textwidth]{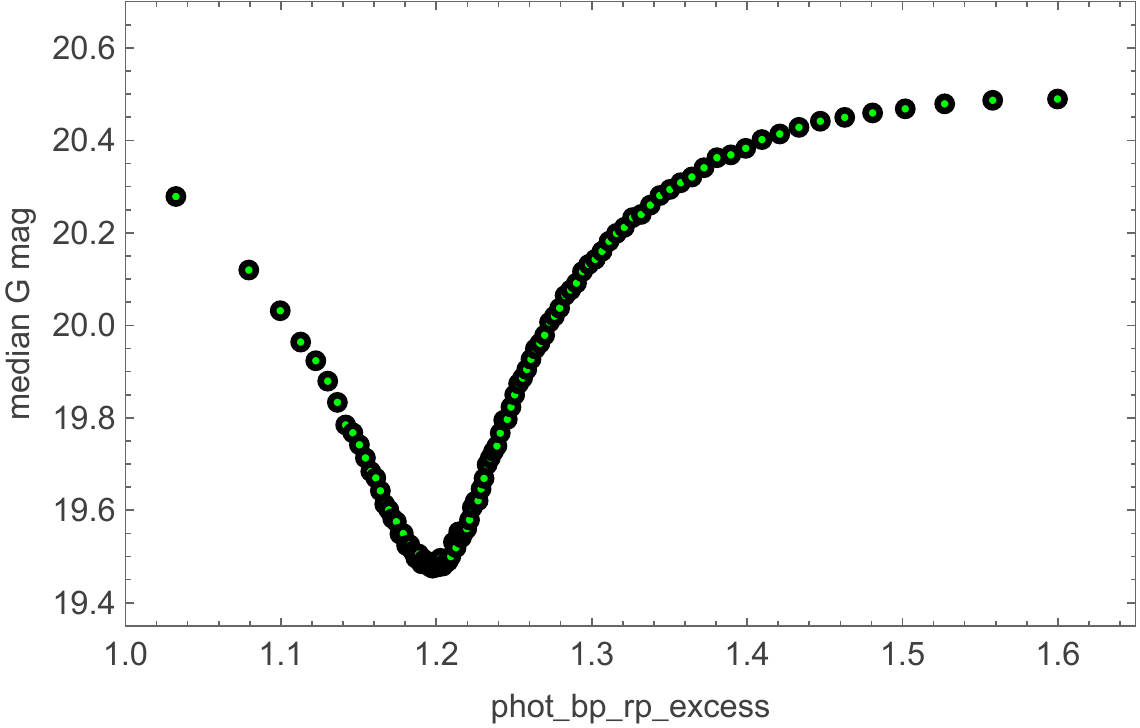}
\caption{
  Median photometric properties of Gaia-CRF3 five-parameter solution sources
  as functions of metadata parameters.
  \emph{Top left:} $G$ magnitude versus right ascension--declination astrometric correlation
  coefficient \texttt{ra\_dec\_corr}. 
  \emph{Top right:} $G$ magnitude versus right ascension--declination error ellipse elongation. 
  \emph{Bottom left:} $\mathrm{BP}-\mathrm{RP}$ color versus \texttt{ra\_dec\_corr}. 
%  \emph{Bottom left:} Median \texttt{phot\_bp\_rp\_excess\_factor} $C$ versus \texttt{ra\_dec\_corr}. 
  \emph{Bottom right:} $G$ magnitude versus \texttt{phot\_bp\_rp\_excess\_factor} $C$. 
  \label{fig:main}
}
\end{figure}

Figure~\ref{fig:main} (top left panel) shows the median $G$ magnitude as a function
of \texttt{ra\_dec\_corr}.
The relation is strikingly symmetric about $\rho(\alpha,\delta) = 0$: sources
with near-zero correlation appear systematically \emph{brighter} (smaller $G$)
than sources at $|\rho(\alpha,\delta)| \approx 0.6$, with a total
peak-to-trough amplitude of approximately 0.4~mag in the median.
The symmetry about zero is a key diagnostic --- the effect responds to
$|\rho(\alpha,\delta)|$, not to its sign. This is further corroborated by a tight monotonic relation of median $G$ versus elongation of the ra-dec error ellipse, which is computed via the ratio of the two covariance eigenvalues (Figure~\ref{fig:main}, top right panel).

A 0.4~mag shift in median $G$ is far too large to be attributed to
photometric calibration zero-point offsets, which are documented at the
mmag to few-hundredths-of-a-magnitude level for EDR3 \citep{Riello2021}.

\subsection{Color versus \texttt{ra\_dec\_corr}}

In contrast, the median $\mathrm{BP} - \mathrm{RP}$ color as a function of
$\rho(\alpha,\delta)$ is markedly asymmetric (Figure~\ref{fig:main}, bottom left
panel). There is no mirror symmetry about zero: positive and negative
$\rho(\alpha,\delta)$ correspond to different sky regions and therefore
sample different large-scale distributions of QSO colors, plausibly reflecting
dust extinction or redshift gradients correlated with Galactic structure. There is a strong north-south asymmetry of the distribution of $\rho(\alpha,\delta)$ in Galactic coordinates.
The range of color variation is $\sim1/20$ of that for $G$ magnitude. Corroborating this, separate plots of BP and RP magnitudes versus $\rho(\alpha,\delta)$ show symmetric curves similar to the $G$ magnitude dependence. This rules out the possibility that the
$G$-magnitude effect is driven by color-dependent photometric biases
(which would themselves follow the asymmetric color distribution).

\subsection{G magnitude  versus BP/RP excess factor}

The \texttt{phot\_bp\_rp\_excess\_factor}, hereafter $C$, measures the ratio of the
summed BP and RP flux to the $G$-band flux and serves as a diagnostic of
photometric consistency \citep{Riello2021}. Elevated values indicate
contamination from nearby sources, extended emission from host galaxies, or internal
photometric inconsistency between the three bands \citep{2023A&A...674A..25H}.

I find that the median $G$ magnitude tightly correlates
with the median $C$ across the full sample as shown in Figure~\ref{fig:main}, bottom right panel, spanning $\sim$1~mag in $G$ over the full range of $C$. 

$C$ as a function of \texttt{ra\_dec\_corr} follows a symmetric curve similar to Figure~\ref{fig:main}, top left panel, with a minimum at $\rho(\alpha,\delta) = 0$, so that
sources with near-zero $\rho(\alpha,\delta)$ display the most internally consistent photometry (smallest $C$), while those at
$|\rho(\alpha,\delta)| \approx 0.6$ show elevated excess factors by up to $0.03$.

\subsection{Astrometric quality: no corresponding RUWE signal}

The Renormalised Unit Weight Error \citep[RUWE;][]{Lindegren2021_astrom}
measures the goodness-of-fit of the astrometric solution and is sensitive
to source multiplicity, extended emission, and astrometric pipeline artifacts.
I find \emph{no} significant dependence of median RUWE on
$\rho(\alpha,\delta)$. The astrometric solution quality is therefore uniform
across the full range of correlation values, confirming that the photometric
anomaly is not accompanied by a corresponding astrometric degradation.
This decoupling rules out a common origin in source crowding or
extended morphology, which would affect both astrometry and photometry.

\subsection{Relative significance of \texttt{phot\_bp\_rp\_excess\_factor} and \texttt{ra\_dec\_corr} for the G-magnitude trend}

To test whether the dependence of median $G$ on $\rho(\alpha,\delta)$
is entirely mediated by \texttt{phot\_bp\_rp\_excess\_factor}, I selected
a narrow slice of 198\,152 five-parameter CRF3 sources with
$1.23 \leq C \leq 1.27$, effectively holding $C$ approximately constant.
Within this restricted sample, the symmetric dependence of median $G$ on
$\rho(\alpha,\delta)$ persists with the same shape and sense as in the
full sample, albeit with increased noise due to the reduced sample size.
The peak-to-trough amplitude is reduced to $\sim$0.3~mag compared to
$\sim$0.4~mag in the full sample.
This result has two implications. First, the $G$--$\rho(\alpha,\delta)$
trend is \emph{not} simply a secondary consequence of the
$G$--$C$ correlation: even at fixed $C$, scan-angle geometry
independently predicts the photometric bias. Second, the modest reduction
in amplitude ($\sim$25\%) when $C$ is controlled indicates that the two
variables share a common component of variance --- consistent with both
being downstream manifestations of the same scan-angle-dependent pipeline
artifact --- while $\rho(\alpha,\delta)$ remains the more fundamental and
stronger driver of the bias. 
%A multivariate correction for systematic $G$-band offsets in CRF3 would therefore need to account for $\rho(\alpha,\delta)$ independently of $C$.

\section{Discussion and Conclusions}

The combination of findings --- (1) a 0.4~mag symmetric dependence of
median $G$ on $|\rho(\alpha,\delta)|$; (2) a correlated and similarly symmetric
minimum in \texttt{phot\_bp\_rp\_excess\_factor}; (3) an asymmetric
$\mathrm{BP}-\mathrm{RP}$ color trend showing that color is a distinct,
sky-position-driven variable; and (4) the absence of any RUWE signal ---
points to a systematic artifact in the Gaia source-detection algorithm that is
tied to the scan-angle distribution encoded in $\rho(\alpha,\delta)$ and error ellipse elongation, rather
than to any intrinsic source property.

Several astrophysical explanations can be excluded. Source isolation and
image compactness are expected to depend on Galactic latitude, not ecliptic latitude;
the two systems are tilted by $\sim 61\degr$, ruling out any such correspondence. The fact that RUWE is uniform
argues against crowding or extended-source effects as the driver.
The expected depth-dependent selection effect — systematically fainter median magnitudes near $\beta=\pm 45\degr$ where transit counts peak — has the opposite sign from the \texttt{ra\_dec\_corr} trend, further confirming the latter is not a completeness artifact.

The similar dependencies of $G$, BP, and RP trends on both \texttt{ra\_dec\_corr} and
\texttt{phot\_bp\_rp\_excess\_factor} strongly argue for a single common upstream cause in the Gaia pipeline.
The 0.4~mag amplitude is large enough to affect photometric completeness
estimates, luminosity function determinations, and selection-function modeling
for surveys built on the CRF3 source list.

%The data underlying this Research Note are available in the Gaia Archive at
%\url{https://gea.esac.esa.int/archive/} via the \texttt{gaiadr3.gaia\_crf3}
%and \texttt{gaiadr3.gaia\_source} tables.

%\begin{acknowledgments}
%This work has made use of data from the European Space Agency (ESA)
%mission \textit{Gaia} (\url{https://www.cosmos.esa.int/gaia}), processed by
%the \textit{Gaia} Data Processing and Analysis Consortium
%(DPAC, \url{https://www.cosmos.esa.int/web/gaia/dpac/consortium}).
%Funding for the DPAC has been provided by national institutions, in particular
%the institutions participating in the \textit{Gaia} Multilateral Agreement.
%\end{acknowledgments}

%\facilities{Gaia}

\bibliography{main}{}
\bibliographystyle{aasjournalv7}

\end{document}